\documentclass[conference]{IEEEtran}
\usepackage{cite}
\usepackage{amsmath,amssymb,amsfonts}
\usepackage{algorithmic}
\usepackage{graphicx}
\usepackage{textcomp}
\usepackage{xcolor}
\usepackage{tcolorbox}
\usepackage{booktabs}
\usepackage{enumitem}
\usepackage{url}
\usepackage{multirow}

\def\BibTeX{{\rm B\kern-.05em{\sc i\kern-.025em b}\kern-.08em
    T\kern-.1667em\lower.7ex\hbox{E}\kern-.125emX}}
\begin{document}

\title{Genre Controlled Music Generation via Activation Steering}

\author{
Swathi Narashiman \quad Pranay Mathur \quad Dipanshu Panda$^{\dagger}$ \quad Jayden Koshy Joe$^{\dagger}$\thanks{Correspondence to: Jayden Koshy Joe (jaydenkjoe@gmail.com).} \quad Harshith M R$^{\dagger}$  \\[1ex]
Anish Veerakumar \quad Aniruddh Krishna \quad Keerthiharan A
\thanks{† Equal contribution. All authors are affiliated with the Indian Institute of Technology Madras.}
}

\maketitle

\begin{abstract}
Computational Music Generation is evolving towards non-conventional styles, demanding
methods that enable precise and controllable blending of diverse music elements. In this
work, we present a method for fine grained control using inference-time interventions on
an autoregressive generative transformer, MusicGen. Through our approach, we achieve
genre control by steering the residual stream using weights of a linear probe on it. By
framing activation steering as a human-controllable interaction, our work highlights how
interpretable model behaviors can empower in co-creative music generation. Audio samples demonstrating our method are available on our demo page \footnote{\url{https://controllable-genre-fusion.github.io/}}. Our implementation and
additional materials are available online \footnote{\url{https://anonymous.4open.science/r/fine-grained-genre-control/}}.
\end{abstract}


\section{Introduction}
\label{sec:intro}

Transformer-based models \cite{vaswani2023attentionneed} have been used to generate music auto-regressively, with a variety of representations like MiDi and compressed audio embeddings being used as input tokens [\cite{huang2018musictransformer}, \cite{shih2022themetransformersymbolicmusic}, \cite{agostinelli2023musiclmgeneratingmusictext}]. Music modeling, with requirements for coherence with regard to harmonic progressions, motifs and musical themes provide a challenging landscape for generative systems. 
Prior work in conditional music generation focuses on music generation guided by text-based, genre or mood descriptions conditioning based on text-based prompts, such as genre or mood descriptions [\cite{lu2023musecocogeneratingsymbolicmusic}, \cite{agostinelli2023musiclmgeneratingmusictext}, \cite{melechovsky2024mustangocontrollabletexttomusicgeneration}]. Subsequent studies also investigated fine-grained control over instruments and chord progressions \cite{tal2024jointaudiosymbolicconditioning}. However, these commonly involve significant compute requirements. 

Considering the challenges involved, we draw inspiration from techniques used to interpret Large Language Models [\cite{arditi2024refusallanguagemodelsmediated}, \cite{li2024inferencetimeinterventionelicitingtruthful}, \cite{turner2024steeringlanguagemodelsactivation}] for a lightweight and effective method to achieve fine-grained control. We analyze MusicGen's \cite{copet2024simplecontrollablemusicgeneration} residual stream and attention layer outputs for music concepts like pitch, instruments, genres and chord progressions. Further, we intervene during inference time to control the generation by adding a steering vector to the residual stream activations. 
\\Our key contributions are:

\begin{itemize}
    \item Providing a user controllable method to steer the activations of MusicGen to enable genre control.
    \item Extending compute efficient activation steering techniques beyond discrete attributes to continuous, non-binary concepts such as musical genres. 
    \item Evaluating our approach through both quantitative (CLAP-based) and human perceptual studies, validating that activation steering enhances genre control while preserving musical coherence.
\end{itemize}

\section{Related Work}
Our contributions extend several prior approaches used to enable style transfer and genre fusion. Timbretron \cite{huang2018musictransformer} introduces a CycleGAN  based approach to affect timbre transfer
while keeping the melodic content the same. However, such an
approach requires training separate models for each pair of instruments. LoRA (Low-Rank Adaptation) \cite{hu2021loralowrankadaptationlarge} suggests an efficient method in comparison to full fine-tuning, however, while fine-tuning is advantageous to learn new concepts, they could hinder performance on other tasks. Conversely, our inference-time interventions can be applied selectively.
\newline
Mechanistic Interpretability aims to understand the internal computational mechanisms in deep neural networks while activation engineering aims to steer behaviour of such deep neural networks without  further training. Several previous work [\cite{arditi2024refusallanguagemodelsmediated}, \cite{turner2024steeringlanguagemodelsactivation}, \cite{li2024inferencetimeinterventionelicitingtruthful}] utilize difference-in-means to steer language models toward desirable behavior e.g. truthfulness. \cite{facchiano2025activationpatchinginterpretablesteering} used activation steering to bring about modulations in measurable concepts like tempo. \textbf{We extend these to scenarios that are not necessarily discrete such as genres.}
\newline
\textbf{SMITIN} \cite{Koo_2025} trains logistic regression probes on MusicGen's attention outputs to add instruments to a given sample. In this work, we extend this idea to genre transfer.

\section{Task Set-Up}
Our goal is to identify the layer at which a particular concept of interest is best linearly represented. For this, we apply k-means clustering to the activations and observe the layer-wise Adjusted Rand Index (ARI) and Normalized Mutual Information (NMI). Since each music sample is passed through MusicGen with its genre known, each activation vector is associated with a true genre label. After clustering, the resulting cluster indices are compared with these labels: ARI measures the extent to which samples of the same genre are grouped into the same cluster (adjusted for chance), while NMI quantifies the mutual dependence between cluster assignments and true labels. For the layer with the highest ARI, we plot the Jaccard similarity matrix and observe significant overlap between the ground-truth labels and clusters obtained, confirming the presence of musical concepts in the MusicGen.
\newline
To observe whether these relationships are linear, we train linear probes on residual stream activations. The probes achieving high accuracy provides support that the concepts of interest are linearly represented, thereby giving a method to disentangle concepts and steer generation.
\newline
\noindent\textbf{Residual Stream Probe:}  
Given residual stream activations at layer $l$ and token position $a$, denoted $\mathbf{h}^l_a$, from datasets $D_0$ (label 0) and $D_1$ (label 1), the probe learns a \textbf{linear mapping}:

\begin{equation}
\hat{y} = W^\top \mathbf{h}^l_a + b
\end{equation}

to predict a concept classifier $\mathcal{G} : \mathcal{X} \to \{0,1\}$:

\begin{equation}
\mathcal{G}(\mathbf{h}^l_a) = 
\begin{cases}
1 & \text{if } \mathbf{h}^l_a \in M(D_1) \\
0 & \text{if } \mathbf{h}^l_a \in M(D_0)
\end{cases}
\end{equation}
where $M(D_i)$ denotes the residual stream activations obtained from the model when evaluated on dataset $D_i \;(i \in {0,1})$. We hypothesize that the layer yielding the highest probe accuracy provides the best linear representation of the concept.

\begin{figure}[h]
  \centering
  \begin{minipage}{0.65\linewidth}
    \includegraphics[width=\linewidth]{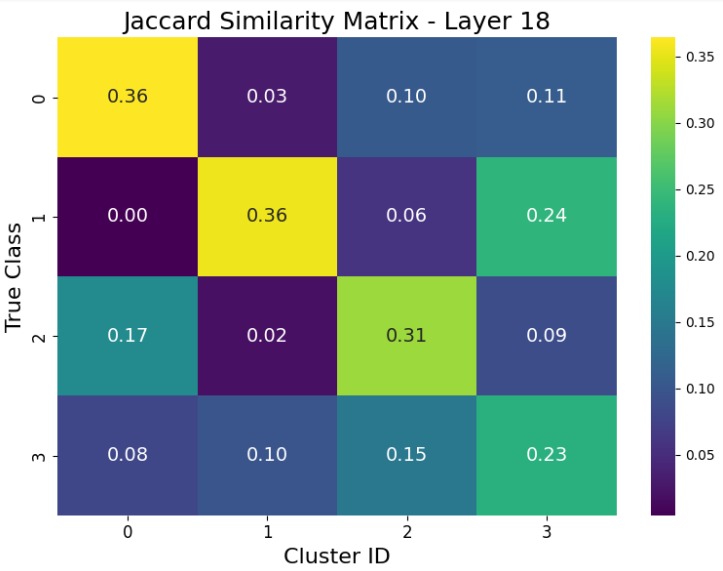}
  \end{minipage}
  \hspace{0.05\linewidth} 
  \begin{minipage}{0.25\linewidth}
    \vspace{-0.5cm} 
    \fbox{%
      \parbox{\linewidth}{
        \textbf{Legend:} \\
        0 - Classical \\
        1 - Electronic \\
        2 - Rock \\
        3 - Jazz
      }%
    }
  \end{minipage}
  \caption{Jaccard Similarity matrix for the genre activations }
  \label{fig:myfigure} 
\end{figure}

\subsection{Background}
We adopt a decoder-only transformer as our generative backbone. Formally, given a vocabulary $\mathcal{V}$ and an input token sequence $\mathbf{x} = (x_1,\dots,x_N)\in\mathcal{V}^N$, the model predicts a distribution over the next token at each position. Concretely, the model outputs a tensor $\mathbf{y}\in\mathbb{R}^{N\times|\mathcal{V}|}$ of logits (or probabilities) for each of the $N$ output positions. At each layer $\ell$, the model maintains a “residual stream” vector $\mathbf{{h}^\ell_i}\in\mathbb{R}^d$ (where $d$ is the hidden dimensionality). Each transformer layer adds contributions from attention and MLP layers to the residual stream as follows:
\begin{equation}
    \Tilde{{h}^{l}_i} = {h}^{l}_i + \texttt{Attn}^{l}({h}^{l}_{1:i})
\end{equation}
\begin{equation}
{h}^{l+1}_i = \Tilde{h}^{l}_i + \texttt{MLP}^{l}(\Tilde{h^{l}_i}) 
\end{equation}

After the final transformer layer, passing it through an unembedding matrix transforms it to a probability distribution over the vocabulary using a softmax activation.
\newline
For music generation we use MusicGen (\cite{copet2024simplecontrollablemusicgeneration}), an auto-regressive transformer operating on Encodec (\cite{défossez2022highfidelityneuralaudio}) audio tokens. MusicGen has four variants: \texttt{small}, \texttt{medium}, \texttt{large}, and \texttt{melody} alongside stereo variants. The first three use cross-attention over T5 text embeddings for conditioning, while \texttt{melody} prepends the conditioning to the input sequence. We perform our experiments on \texttt{musicgen-small}, and our techniques can be extended to \texttt{musicgen-medium} and \texttt{musicgen-large} directly.

\subsection{Genre probing}
A subset of samples from \texttt{lewtun/music\_genres} \footnote{\url{https://huggingface.co/datasets/lewtun/music\_genres}}
 dataset (\cite{lewtun2023music}) is used in this work. The original music genres dataset is a large audio classification corpus hosted on the Hugging Face Datasets platform, containing approximately 25000 labeled music clips with associated metadata (raw audio, a unique song identifier, and genre labels) organized into standard train and test splits. Each example includes an audio waveform and genre annotation, supporting a multi-class taxonomy of around 19 distinct genres, with clip duration of 30 s. From this corpus, we extract fixed-length 10-second segments filtered to four genre categories: \texttt{rock}, \texttt{classical}, \texttt{electronic}, and \texttt{jazz}, resulting in a balanced dataset of 900 samples per class for downstream experiments.
\newline
Since our objective involves classification, we attempt training on the a collection of $n$ genres represented with one-hot encoding for multi-label classification.  We collected the residual stream activations at all layers for the samples using a single A100 GPU for a period of 1.5 hours. 

\subsection{Extracting Feature Vectors via Linear Probe Weights}

Inspired by literature on interpreting neural networks with the use of probes, we train linear classifiers on the residual stream activations to predict the target label. Specifically, for each residual stream vector at a particular layer $\mathbf{h^l_a}$, we predict the label associated with each audio sample using:
\begin{equation}
    \hat{y^l_a} = W_l^T \mathbf{h^l_a} + b^l_a
\end{equation}
We train the probe while keeping the weights of the transformer frozen, using mean-squared error loss. In effect, the weights of this probe learns a linear direction that best separates the two classes, and hence, we use it to steer generation.
\subsection{Inference-time Intervention}
Given a unit-magnitude steering vector $\mathbf{r^l}$, we add a scaled version to the residual stream before the layer with the best linear representation of the concept and then normalize to unit norm.
\begin{equation}
    \mathbf{h^l_i} \leftarrow \frac{\mathbf{h^l_i} + \alpha \mathbf{r^l}}{1 + \alpha}
\end{equation}
We performed a hyperparameter search over values of $\alpha$ in the range $[0.3, 1.0]$, 
where $\alpha$ is a scaling factor that controls the strength of the intervention in 
activation steering. Interventions are made only to the conditional residual stream.

\section{Experiments}

We assess our approach by generating audio samples that smoothly transition between genres, namely \texttt{classical} → \texttt{electronic}, \texttt{electronic} → \texttt{jazz}, \texttt{jazz} → \texttt{rock}, and \texttt{rock} → \texttt{classical}, followed by quantitative and qualitative evaluations. This pattern was chosen to ensure that for each genre, one experiment steered from it, and one experiment steered to it. If both are successful, we can confidently state that steering can occur between any of the two genres.

\subsection{Prompting Genre Transitions}
To generate genre transitions using text prompting alone (without activation steering), we provide genre descriptive prompts to the model corresponding to each source–target genre pair. The prompts were designed \footnote{This prompt template was taken from the official musicgen demo page for various music samples: \url{https://ai.honu.io/papers/musicgen/}} to reflect the target genre’s musical characteristics while maintaining generality across instruments and style.

\begin{tcolorbox}[colback=gray!5,colframe=gray!80,title=\textbf{Genre Transition Prompts}]
\itshape
\begin{itemize}[leftmargin=1pt, itemsep=0pt, topsep=0pt, parsep=0pt, partopsep=0pt]
    \item \textbf{Rock $\rightarrow$ Classical:} ``An orchestra playing a classical piece, with a strong brass section and strings.''
    \item \textbf{Classical $\rightarrow$ Electronic:} ``An upbeat electronic track with a catchy beat.''
    \item \textbf{Electronic $\rightarrow$ Jazz:} ``Smooth jazz with a saxophone solo and calm piano.''
    \item \textbf{Jazz $\rightarrow$ Rock:} ``Rock with saturated guitars, a heavy bass line, and crazy drum breaks and fills.''
\end{itemize}
\end{tcolorbox}

These text prompts were passed to the generative model as conditioning inputs. The corresponding \textit{steered} versions used the same input audio without text conditioning and the transitions were effected entirely by interventions on the activations. The resulting prompted outputs serve as a baseline for evaluating the perceptual and semantic improvements introduced by activation steering.

Steering effects a genre transition while maintaining certain features of the original audio, like tempo and the melodic line. We hypothesize this to different features being linearly represented at varying layers of the transformer.

\subsection{Quantitative evaluation setup}
We evaluate our intervention method using \texttt{clap-htsat-unfused} model. We utilized the zero-shot
audio classification pipeline available via HuggingFace. 

\noindent To evaluate the success of genre transition:
\begin{enumerate}
    \item Let $x_{\text{orig}}$ be the original audio generated using a prompt from the \textbf{source genre}.
    \item Let $x_{\text{steered}}$ be the audio obtained after steering the model toward the \textbf{target genre}.
    \item We employ the CLAP model in a zero-shot classification setting with the set of genre labels $\mathcal{G} = \{g_1, g_2\}$ that includes both the source and target genres.
    \item For each audio sample $x$, we compute:
    \[
    \text{CLAP}(x, g) \in [0, 1]
    \]
    which denotes the similarity score between the audio and the genre label $g$ as interpreted by the CLAP model.
     \item The difference of the score provided to $x_{steered}$ and $x_{orig}$ is evaluated to determine the extent of genre transition.
\end{enumerate}

\subsection{Qualitative listening study}

We conduct a controlled perceptual listening study to assess the perceptual quality and genre-blending behavior of the proposed genre transition method. While this evaluation provides subjective insights into listener preferences, we emphasize that it is intended as a preliminary perceptual assessment rather than a comprehensive human-centric evaluation.

The study follows a two-alternative forced choice (2AFC) paradigm and involves 24 participants recruited via university mailing lists and peer networks. Among them, 7 participants were trained musicians with formal musical education ($\geq$3 years of training), while the remaining 17 participants were non-musicians without formal music training. This composition was chosen to capture both expert and lay listener perspectives, reflecting real-world listening scenarios where music generation systems are primarily consumed by non-expert users.

Each participant completed 24 randomized trials, with each trial consisting of a pair of 10-second audio samples. The trials covered four genre transition pairs: \texttt{classical}→\texttt{electronic}, \texttt{electronic}→\texttt{jazz}, \texttt{jazz}→\texttt{rock}, and \texttt{rock}→\texttt{classical}. 

For each genre transition, six audio pairs were randomly sampled from a pool of 20 unique source samples. Each pair consisted of (i) a prompted baseline, generated using only textual instructions to combine source and target genres, and (ii) a steered sample generated using the proposed method. The order of presentation (A/B) was randomized across trials to mitigate ordering bias.

During each trial, participants were instructed to listen to both samples in full and select the one that better achieved a perceptually coherent blend of the two target genres. No time constraints were imposed, and participants were allowed to replay samples before making their selection.

\begin{table}[!t]
\caption{Mean CLAP scores and change (Modified $-$ Original) for steered vs.\ prompted transitions. All CLAP scores refer to the target genre. In multiple cases, steering results in larger changes in CLAP scores at comparable variances compared to prompting.}
\label{tab:clap}

\setlength{\tabcolsep}{4pt}
\footnotesize
\resizebox{\columnwidth}{!}{%
\begin{tabular}{lccc}
\toprule
\textbf{Transition} &
\textbf{Original} &
\textbf{Modified} &
\textbf{$\Delta$ CLAP} \\
\midrule
Rock $\rightarrow$ Classical
  & $0.0703\!\pm\!0.0685$
  & $0.1785\!\pm\!0.1270$
  & $\mathbf{+0.1081}$ (Steered) \\
  & $0.0888\!\pm\!0.1074$
  & $0.1851\!\pm\!0.1223$
  & $+0.0963$ (Prompted) \\
\midrule
Classical $\rightarrow$ Electronic
  & $0.0252\!\pm\!0.0164$
  & $0.2309\!\pm\!0.1329$
  & $\mathbf{+0.2057}$ (Steered) \\
  & $0.0267\!\pm\!0.0158$
  & $0.1542\!\pm\!0.1484$
  & $+0.1276$ (Prompted) \\
\midrule
Electronic $\rightarrow$ Jazz
  & $0.3398\!\pm\!0.1153$
  & $0.3038\!\pm\!0.1150$
  & $-0.0359$ (Steered) \\
  & $0.3517\!\pm\!0.1224$
  & $0.2639\!\pm\!0.1427$
  & $-0.0878$ (Prompted) \\
\midrule
Jazz $\rightarrow$ Rock
  & $0.0009\!\pm\!0.0005$
  & $0.1524\!\pm\!0.1641$
  & $\mathbf{+0.1515}$ (Steered) \\
  & $0.0009\!\pm\!0.0005$
  & $0.0475\!\pm\!0.0747$
  & $+0.0466$ (Prompted) \\
\bottomrule
\end{tabular}
}
\end{table}

\section{Results}

Table~\ref{tab:clap} summarizes the CLAP score changes obtained under steered and prompted genre transitions.

Our CLAP scores verify the improved adherence to target class labels after steering over prompting. Steering consistently produces more pronounced improvements in CLAP scores than prompting, even at comparable variances, highlighting the effectiveness of activation-based control. 

\begin{table}[h!]
\centering
\caption{2AFC evaluation results across genre transitions for practiced musicians and general audience. Accuracy here refers to the preference to steered samples over prompted samples. P-values are obtained using a binomial test against chance (0.5)}
\begin{tabular}{@{}l l c c@{}}
\toprule
\textbf{Participant Group} & \textbf{Transition} & \textbf{Accuracy} & \textbf{P-value} \\ 
\midrule

\multirow{4}{*}{\textbf{Practiced Musicians}} 
& Rock $\rightarrow$ Classical & 0.805 & 0.0001 \\
& Classical $\rightarrow$ Electronic & 0.738 & 0.0028 \\
& Electronic $\rightarrow$ Jazz & 0.785 & 0.0003 \\
& Jazz $\rightarrow$ Rock & 0.805 & 0.0001 \\

\midrule

\multirow{4}{*}{\textbf{General Audience}} 
& Rock $\rightarrow$ Classical & 0.726 & 0.0001 \\
& Classical $\rightarrow$ Electronic & 0.595 & 0.0690 \\
& Electronic $\rightarrow$ Jazz & 0.696 & 0.0001 \\
& Jazz $\rightarrow$ Rock & 0.700 & 0.0001 \\

\bottomrule
\end{tabular}
\end{table}

The 2AFC results (Table 2) indicate that both practiced musicians and general listeners consistently preferred steered samples across all genre transitions, with higher accuracy and statistical significance in most cases. This aligns with the CLAP-based quantitative evaluation (Table 1), where steering generally produces larger positive shifts in target-genre similarity compared to prompting. Together, these findings confirm that activation steering offers perceptually and semantically stronger control over genre transition than mere textual prompting.

\section{Conclusion}
In this work, we achieve enhanced genre control with a lightweight intervention method. By extending Activation Engineering techniques from LLMs, we demonstrate effective manipulation of musical features even in the absence of clearly contrasting samples. Our qualitative study further validates the perceptual effectiveness of the approach. Both trained musicians and casual listeners consistently preferred steered samples across genre transitions.

While CLAP scores provide a useful quantitative proxy for genre alignment, they may not directly represent genre semantics. Instead, they likely capture correlated musical attributes (such as timbre, tempo, rhythm)
that align with genre distinctions, thereby supporting our approach. Because our work is intended for human-centric creative workflows, the steering strength functions as an interactive, user-controllable parameter rather than a fixed hyperparameter. Future work could explore principled mappings between perceptual preferences and steering magnitudes, as well as deeper analyses of superposition phenomena in attention and residual stream activations. Such insights may enable more generalizable and interpretable intervention strategies for controllable music generation.

\bibliographystyle{IEEEtran}
\bibliography{icme2026references}
\end{document}